\documentstyle[12pt]{article}
\newcommand{\beq}{\begin{equation}}
\newcommand{\eeq}{\end{equation}}

\begin{document}
\begin{titlepage}
\begin{flushright}
NBI-HE-94-21\\
March 1994\\
\end{flushright}
\vspace{0.5cm}
\begin{center}
{\large {\bf Cosmic String Nucleation near the Inflationary Phase Boundary}}\\
\vspace{1.5cm}

{\bf Minos Axenides}
\footnote{e-mail: axenides@nbivax.nbi.dk}\\
\vspace{0.4cm}
{\em The Niels Bohr Institute, University of Copenhagen\\
and\\
Theoretical Astrophysics Center\\
17 Blegdamsvej, 2100 Copenhagen \O, Denmark}\\
\vspace{1cm}
{\bf Arne L. Larsen}
\footnote{e-mail: allarsen@nbivax.nbi.dk}\\
\vspace{0.4cm}
{\em Nordita\\
17 Blegdamsvej, 2100 Copenhagen \O, Denmark} \\

\vspace{0.4cm}
\end{center}
\vspace{6mm}
\begin{abstract}
\hspace*{-6mm}We investigate
the nucleation of circular cosmic strings in models of
generalized inflationary universes with an accelerating scale factor. We
consider toy cosmological models of a smooth
inflationary exit and transition into a flat Minkowski spacetime. Our
results establish that an inflationary expanding phase is necessary but
not sufficient for quantum nucleation of circular cosmic strings
to occur.
\end{abstract}
\end{titlepage}
\newpage

\section{Introduction and Conclusions}
Inflation \cite{Linde,Rocky,RB} is a
short period of rapid expansion in the early
history of the universe whereby its presently observable part
originated from a tiny initial region. Topological defects such as
cosmic strings, monopoles, and domain walls  are extended objects
present in the spectrum of grand unified theories that are believed to
be typically generated in phase transitions in the early universe.
They could have acted as seeds for the generation of density
perturbations that resulted in the large scale structure of the universe
and the observed anisotropy in the cosmic microwave background radiation.
Recently it was realized that such objects can be created spontaneously
in a  de Sitter spacetime through the process of quantum
nucleation \cite{Guth}. More specifically, using the static
parametrization it was found that the classical evolution of a circular
string is determined by a simple potential barrier and that strings can
nucleate by a quantum mechanical tunnelling through the barrier. In a
realistic situation any topological defects formed at the onset of the
inflationary period are expected to be inflated away. Strings nucleated
towards the inflationary exit will eventually contract upon their
entrance into the radiation dominated phase inside the causal
horizon \cite{JG}.
If they are still circular they are expected to form black
holes \cite{SH}, otherwise they oscillate radiating away their
gravitational energy. In fact this was the conclusion in a cosmological
model whereby a de Sitter phase is followed "abruptly"  by a flat
Minkowskian spacetime at which point the Hubble rate of expansion
changes discontinuously \cite{JG}.

It is
the purpose of our present work to explore how a smooth and
continuous exit from the inflationary era and the passage of the
expanding universe into a radiation dominated one relates to the switch
off of the quantum nucleation of circular cosmic strings. This is of
some phenomenological importance as the actual size distribution of
string loops upon their entrance within our causal horizon must depend
on the precise way in which their nucleation ended as the
exponential expansion of the universe
relaxed into its radiation dominated rate. This
problem also hinges on the still unresolved issue of the actual
realization of inflation in the cosmic history. A number of models have
indeed been proposed which correspond to cosmologies with accelerating
scale factors ($d^2 a/dt^2>0$) and variable cosmic time
dependance $a(t)$. In a previous work \cite{arne} we took a first step towards
the important issue of how generic is cosmic string nucleation in
inflationary cosmologies. We specifically studied circular string evolution
in a time dependent spacetime ($a(t)\simeq t^p;\; p>0$) with no
static parametrization. In cosmological spacetimes which admit circular
string configurations with the property:
\begin{equation}
\exists\; t_{0} :\;\; f(t) \geq f(t_{0}) > 0 ,\;\;\;\;\;\;  \frac{df}
{dt}(t_{0})=0,
\end{equation}
where $f$ is the physical radius of the loop and $t$ is the cosmic time,
it is reasonable to assume that string nucleation plays an important
role. The presence of $t_{0}$ in the time dependent spacetime has of course
only a local significance. It implies that a contracting circular string
configuration hits a potential barrier and is energetically prohibited to
collapse. The existence of such string loops is a necessary condition
for string tunnelling and nucleation in the spacetime under
consideration. In fact we found that for power law expanding spacetimes
with $p< 2\sqrt{2}+3$ no such configurations exist. This we interpreted
as the absence of string tunnelling. It followed both from an explicit
numerical search for such solutions to the string equations of motion
and from the analytical observation that eq.(1.1) in fact implies
$\ddot{f}(t_0) \geq 0$, where a dot denotes differentiation with respect to
$t$. In conjuction with the derived string equations of
motion it implies that the Hubble rate of expansion $ H(t)=
\dot{a}(t)/{a(t)}$ at $t=t_{0}$ satisfies a
master inequality:
\begin{equation}
\dot{H}(t_{0}) \geq (2\sqrt{2}-3) H(t_0)^2
\end{equation}
It provides us with a necessary condition in order that a circular
string nucleates at $t=t_{0}$ with a finite radius $f(t_{0})$ by tunnelling
through a barrier. Loosely speaking we may say that for time dependent
spacetimes a potential barrier that prevents circular strings from
collapsing at $t=t_{0}$ will change in time and in some cases it may only
exist for a finite amount of time during the evolution of the universe.
This implies that a string nucleated at one moment may collapse at a
later one by classically shrinking to a point in the absence of any
barrier. This is a very realistic and physical picture as our universe
went through a de Sitter like expansion for a short period in its cosmic
history before it entered the radiation dominated era where the string
evolution equations indicate an unconditional collapse for circular
strings. The precise way in which the Hubble expansion relaxed to its
radiation dominated era is not known. Under the hypothesis that the
scale factor traversed smoothly through all possible time behaviours before it
relaxed to its present one, we investigate cosmic string nucleation in
models with generalized inflation ($\ddot{a}(t)>0$). More
specifically we study the cases of intermediate and superinflation
\cite{JB}. Both our analytical and numerical tools suggest that
circular string nucleation is almost generic. For a given model of
inflationary cosmology we precisely determine the time at which
nucleation either sets in or switches off. Interestingly we find that in
contrast to the purely de Sitter case, where $H(t)$ is constant, string
nucleation
does not occur throughout the inflationary phase. Furthermore we proceed
and present a toy cosmological model whereby the universe after
exponentially expanding for a period of time smoothly
enters a flat Minkowskian phase. We find that circular string nucleation
stops before the universe exits from the inflationary era.

In conclusion, in cosmological models with generalized inflation
cosmic string nucleation appears to be almost
generic in the sense that inflation is a necessary but not sufficient
condition. As a consequence, for a cosmologically smooth transition from
an inflationary era with $\ddot{a}(t)>0$ and into a radiation
dominated one $(\ddot{a}(t)<0)$, string nucleation ceases in a
well-defined and determined manner at some $t=t_{0}$, which
nevertheless is model dependent. Circular loops eventually collapse and
form black holes upon entering the causal horizon whereas noncircular
ones are likely to become the seeds of density perturbations. Their
precise size distribution is of interest for models of large and medium
scale structure formation in the late universe.

The
paper is organized as follows: In Section 2 we derive the circular
string evolution in an arbitrary spatially
flat FRW spacetime. In Section 3 we investigate
string nucleation for models which exhibit generalized inflation
($\ddot{a}(t)>0$) and present a toy model for a smooth transition
out of a de Sitter phase and into a flat Minkowskian one.

\section{String Evolution Equations}
\setcounter{equation}{0}
We want to derive the equations of motion for circular strings in a
spatially flat Friedman-Robertson-Walker (FRW) spacetime. These
spacetimes are usually parametrized in terms of comoving coordinates:
\begin{equation}
ds^{2} = -dt^{2} + a(t)^{2}\left(
dr^{2}+r^{2}d\theta^{2}+r^{2}\sin^{2}\theta d\phi^2\right),
\end{equation}
where $a(t)$ is the scale factor. For an observer who employs the
comoving coordinate system the circular string will appear to have a
physical radius $f$ at the cosmic time $t$:
\begin{equation}
f(t)\;=\;r\;a(t).
\end{equation}
It is convenient for our purposes to use a parametrization for the
spacetime directly in terms of these variables. From eqs.(2.1) and (2.2)
it is easy to obtain:
\begin{equation}
ds^{2}=-(1-H^{2}f^{2})dt^{2}-2Hf\;dfdt + df^{2} +
f^{2}(d\theta^{2}+\sin^2\theta d\phi^{2}).
\end{equation}
Here we have introduced the Hubble expansion rate:
\begin{equation}
H\;=\;\frac{1}{a}\frac{da}{dt}.
\end{equation}
It is generically a function of the cosmic time $t$. Indeed for all
generalized inflationary models with $\ddot{a}(t)>0$ this is
precisely the case with the exception of de Sitter space
$(H=const.\equiv H_0\neq 0)$. In
the latter case we notice that the spacetime metric
in eq.(2.3) admits a stationary but non-static parametrization. It is,
moreover, related to the well-known static parametrization by the
transformation:
\begin{equation}
t_{static}\;=\;t\;-\;\frac{\log\mid 1-H_{0}^2f^{2}\mid}{2H_{0}}
\end{equation}
In what follows, however, we will be mostly interested in spacetimes
with time dependent Hubble expansion rate.

A family of circular strings with a time dependent radius is obtained by
the ansatz:
\begin{equation}
t=\tau,\;\;f=f(t),\;\;\theta=\frac{\pi}{2},\;\;\phi=\sigma.
\end{equation}
Here we have identified the cosmic time with the world-sheet time, and
the azimuthal angle with the world-sheet spatial and periodic
coordinate. The radius $f(t)$ is to be determined by the equations of
motion that depend on the specific string model. We hereby consider the
Nambu-Goto action:

\begin{equation}
{\cal S}= \int\;d\tau d\sigma\; \sqrt {-\det G_{\alpha\beta}},
\end{equation}
where $G_{\alpha\beta}$ is the induced metric on the world-sheet:
\begin{equation}
G_{\alpha\beta}\;=\;g_{\mu\nu}\;X^{\mu}_{,\alpha}\;X^{\nu}_{,\beta}
\end{equation}
with  $X^{\mu}=(t,r,\theta,\phi)$ and
$g_{\mu\nu}$ being given by eq.(2.3).
By taking also into account the ansatz of eq.(2.6) we obtain:
\begin{eqnarray}
&G_{\tau\tau}\;=\;\dot{f}^{2}-2Hf\dot{f}+H^{2}f^{2}-1,\;\;G_{\sigma\sigma}=
f^{2},\;\;G_{\tau\sigma}=0,&\nonumber\\
&\sqrt {-\det G_{\alpha\beta}}\;=\;f\sqrt
{1-(\dot{f}-Hf)^2}\;&
\end{eqnarray}
and we require for a time-like string:
\begin{equation}
(\dot{f}- H f)^2 < 1.
\end{equation}
It is now straightforward to derive the equation of motion \cite{arne}:

\begin{eqnarray}
\ddot{f}f\;-\;2Hf\dot{f}^{3\hspace*{-1mm}}&+&\hspace*{-1mm}(6H^{2}f^{2}-1)
\dot{f}^{2}\;+\;3Hf(1-2H^{2}f^{2})\dot{f}\nonumber\\
\hspace*{-1mm}&-&\hspace*{-1mm}f^{2}\dot{H}\;+\;2H^{4}
f^{4}\;-3H^{2}f^{2}\;+\;1=\;0.
\end{eqnarray}
It determines the physical string size as a function of the cosmic time
in a spatially flat FRW spacetime with a Hubble rate of expansion H. The
complete analytical solution to this equation is unfortunately not
known. It is still though possible to extract some information about the
evolution of the circular strings. This was already discussed in our
previous work \cite{arne}. We nevertheless highlight once more the main
points. A contracting string $(\dot{f}<0)$ will continue to contract
unless there exists a critical time $t_{0}$ such that:
\begin{equation}
f(t)\;\geq\;f(t_0)\;>\;0, \;\;\;\;\;\;\;\; \dot{f}(t_0)=0.
\end{equation}
This should hold true in a time dependent spacetime locally around some
$t=t_0$. The existence of a $t_0$ that satisfies eq.(2.12) means that
the contracting string hits a barrier and is therefore energetically
forbidden to collapse. Such a bounce is a necessary condition for the
existence of a potential barrier and consequently of string tunnelling
and nucleation in the spacetime under consideration. Such an
interpretation is consistent with the results obtained already
\cite{Guth} for the case of quantum creation of circular strings in
de Sitter spacetime. A more detailed discussion for the case of an
arbitrary scale factor can be found in \cite{arne}.

By a Taylor expansion it follows that
$\ddot{f}(t_0)\geq 0$ and then eqs.(2.10)-(2.11) lead to \cite{arne}:
\begin{equation}
\dot{H}(t_0)\;\;\geq\;\;(2\sqrt{2}-3)\;H(t_0)^{2}.
\end{equation}
This is our master inequality which provides a necessary condition for a
cosmic string to nucleate with a finite radius $f(t_0)$ at a time $t_0$
after tunnelling through the barrier. In the limiting case of
equality eq.(2.13) is readily solved by:
\begin{equation}
a(t)\;\;\propto\;\;t^{2\sqrt{2}+3}.
\end{equation}
For a power law inflationary universe
($a(t)\propto t^{p}\;$) with $\;p\geq\;2\sqrt{2}+3$ inequality (2.13) is
fulfilled and string nucleation is expected for any $t_0\in [0,\infty[\;$,
whereas for $p<2\sqrt{2}+3$ no nucleation is expected to occur.
Notice that power law expanding universes have a special property, namely that
$\dot{H}$ is proportional to $H^{2}.$ This implies that any dependence on
the cosmic time $t_0$ drops out from (2.13) with the power of
expansion being the only parameter left.

\section{String Nucleation and Generalized Inflation}
\setcounter{equation}{0}
In this section we proceed to employ both our analytical master
inequality (2.13) and string evolution equation (2.11) to
investigate circular string
nucleation in a wide class of inflationary models. Such a diversity of
spacetimes was recently shown to arise from the equation of state
\cite{JB}:
\begin{equation}
p\;+\;\rho\;=\;\gamma\rho^{h}.
\end{equation}
Here $p$ and $\rho$ are the matter pressure and density whereas $\gamma$ and
$h$ are arbitrary constants. The resulting spatially flat FRW universes
are described by scale factors of the form:
\begin{equation}
a(t)\;\simeq\;e^{\pm t^p},\;\;\;\; e^{\pm e^{\pm t}},\;\;\;\; t^p.
\end{equation}
The question of circular string tunnelling and nucleation in the power
law expanding universes ($a(t)\propto t^p$) has
already been dealt with \cite{arne}.
It was found that a necessary condition for nucleation is that
$p\;\geq\;2\sqrt{2}+3$. This is true for all $t_0 \in [0,\infty[\;$. In the
remaining of this section we show that the latter statement does not
hold for the other types of cosmologies given by eq.(3.2). In order to
be more specific we will first consider the scale factors of the form:
\begin{equation}
a(t)\;=\;e^{a_pt^p};\;\;\;      t\geq 0,
\end{equation}
where $a_p$ is a dimensionful constant. This family of cosmologies
includes de Sitter space $(p=1)$ as a special case. In what follows we
will be only interested in expanding universes $(a_p>0,\;p>0)$ and
$(a_p<0,\;p<0)$, respectively.
\vskip 6pt
\hspace*{-6mm}I. $a_p>0,$ $p>0.$
The necessary condition for string nucleation (2.13) implies:
\begin{equation}
p - 1\;\geq\;\;(2\sqrt{2}-3)\;p\;a_p\;t_{0}^{p}.
\end{equation}
We may observe that for $p\geq 1$ the above
inequality is always fulfilled independently of the nucleation time
$t_0$. This is to be interpreted that we should expect strings to
nucleate at any time $t_0 \in [0,\infty[\;$ of the inflationary phase.
In the special case of $p=1$ this is consistent with the conclusions of
Basu, Guth and Vilenkin \cite{Guth}. It is
not surprising that the de Sitter $(p=1)$ result generalizes to the
superinflationary cosmologies $(p>1)$. Indeed the master inequality (2.13)
expresses the fact that a certain amount of inflation is necessary for
nucleation of circular strings. In this case the superinflationary
universe undergoes a faster expansion $\dot{H} > 0$ than the de Sitter
space $(\dot{H}=0)$.

In the case of $p\in\; ]0,1[\;$, which corresponds to
models of the so-called intermediate inflation type \cite{JB},
inequality (2.13) leads to:
\begin{equation}
a_p^{1/p}\;t_0\;\geq\; (3+2\sqrt{2})^{1/p} (\frac{1}{p}-1)^{1/p}.
\end{equation}
This is a completely different result from the one obtained for $p\geq 1$
and from that of the power law inflationary universes. The inequality (2.13)
gives us now a critical time $t_0^{crit}$ with the interpretation that
nucleation can only take place during the $t\geq t_{0}^{crit}$ period in
the evolution of the universe. In the intermediate inflationary models
there is actually only inflation $(\ddot{a}(t)>0)$ for $t>t^{inf}$
where:
\begin{equation}
a_{p}^{1/p}\;t^{inf}\;=\;(\frac{1}{p}-1)^{1/p}.
\end{equation}
It is interesting to compare the two time scales $t_{0}^{crit}$
and $t^{inf}$ for
different cosmological models ($p$ values). This is done in Fig.1. We
observe that for $p\rightarrow 1$ both
nucleation and inflation start at $t\approx 0$.
For $p\rightarrow 0$ inflation
starts long before nucleation. This is certainly of
little observational value from the point of view of our presently
radiation dominated universe. It demonstrates nevertheless the fact that
inflation is necessary but not sufficient for the nucleation phenomenon
of circular cosmic strings to occur.
\vskip 6pt
\hspace*{-6mm}II. $a_p<0,$ $p<0.$
The cosmologies represented by this class of models are different from
case I. They start at $t=0$ with $a=0$, run through a transient
inflationary era and finally enter into a stable static Minkowski space
$a(\infty)=1$. The inequality (2.13) now takes the form:
\begin{equation}
|a_p|^{-1/|p|}\;t_0\;\;\leq(3-2\sqrt{2})^{1/|p|}
(1+\frac{1}{|p|})^{-1/|p|}.
\end{equation}
The presence of a critical time $t_0^{crit}$ now represents the point
at which nucleation ceases to occur. In this sense it is the endpoint of
the nucleating phase $(t\leq t_0^{crit})$. This timescale is to be compared
with that of the exit from the inflationary phase $(t<t^{inf})$ where
$t^{inf}$ is given by:
\begin{equation}
|a_p|^{-1/|p|}\;t^{inf}\;=\;(1\;+\;\frac{1}
{|p|})^{-1/|p|}.
\end{equation}
In Fig.2 we give the two timescales for different cosmological models.
For $p\rightarrow -\infty$  we may observe that circular string nucleation is
allowed for the entire inflationary era while for $p\approx-1.6$ the
available time for inflationary stretching of the nucleated strings is
maximal. This is the physical picture for cosmological models where a de
Sitter phase undergoes a smooth transition into the radiation dominated
one.

Let us now turn to the second family of scale factors (eq.(3.2)), obtained
by Barrow \cite{JB}, of the form:
\begin{equation}
a(t)\;=\;e^{A e^{Bt}};  \;\;\;\;\; t\in\;]-\infty,\infty[,
\end{equation}
where $A$ and $B$ are constants. We are again only interested in expanding
universes. We distinguish two cases of interest ($A>0,\;B>0$) and
($A<0,\;B<0$).
The first case is somewhat similar to the superinflationary models
already discussed after eq.(3.4). As the expansion rate is now even faster
inequality (2.13) is trivially fulfilled and we expect nucleation to
occur for any $t_0\in\;]-\infty,\infty[\;.$

For the second case inequality (2.13) gives:
\begin{equation}
t_0\;\;\leq\;\;\frac{1}{B}\;\log\;[- \frac{(3+2\sqrt{2})}{A}]
\end{equation}
while inflation takes place for  $t<t^{inf},$ where:
\begin{equation}
t^{inf}\;=\; \frac{1}{B}\;\log\;(-\frac{1}{A})
\end{equation}
It follows that:
\begin{equation}
t^{inf}\;-\;t_{0}\;\;\geq\;\;-\frac{1}{B}\;\;\log\;[3+2\sqrt{2}]
\end{equation}
As $B < 0$ we have the string nucleation switched off to predate the
inflationary exit. We may consider as a specific example the case
$A=B=-1$. The scale factor is shown in Fig.3. The universe inflates for
$t\in\;]-\infty,0[\;$, but there can only be nucleation for
$t\in\;]-\infty\;,\;-\log(3+2\sqrt{2})].$ Therefore, any nucleated
circular string will be stretched by inflation for at least $\Delta
t\;=\;\log (3+2\sqrt{2})$ amount of time in this model. After the
inflationary era the universe smoothly undergoes a transition into
flat Minkowski space, where the circular strings eventually collapse.

We would like finally to mention another toy model that also illustrates a
smooth exit out of a de Sitter inflationary phase and into a flat
Minkowski spacetime. Its scale factor is given by:
\begin{equation}
a(t)\;=\;\frac{e^{H_{0}t/2}}{e^{H_{0}t/2} + e^{-H_{0}t/2}}
\end{equation}
It is evident that we can identify a de Sitter phase for $t\rightarrow -\infty$
$\;(a(t)=e^{H_{0}t})$ and a flat Minkowski one for
$t\rightarrow \infty\;(a(t)=1$) with a smooth interpolation between them. The
relevant quantities for string nucleation are given by:
\begin{eqnarray}
H(t)\;&=&\; \frac{H_{0}}{2}\;[1\;-\;\tanh(H_{0}t/2)],\\
\dot{H}(t)\;&=&\;-\frac{H_{0}^2}{4\cosh^2(H_{0}t/2)}.
\end{eqnarray}
Inequality (2.13) implies that at $H_{0}t_{0}^{crit}
=\log(3-2\sqrt{2})\approx -1.76$ string nucleation ceases
to occur. By evaluating $\ddot{a}(t)$ we may conclude that
the inflationary exit occurs at $H_{0}t^{inf}=0$.
Indeed explicit numerical solutions of the string evolution in eq.(2.11),
depicted in Fig.4, confirm our analytical expectations. Contracting
circular strings can hit a barrier for $H_{0}t\leq H_{0}t_{0}^{crit} \approx
-1.76$ only, after which they expand. At their entrance into the Minkowski
phase $(t>>0)$ the strings collapse.

\newpage
\subsection*{Figure Captions}
\vskip 6pt
Fig.1. Cosmic string nucleation $(t\geq t_0^{crit})$ near the inflationary
boundary $t>t^{inf}$ in cosmologies with scale factors
$a(t)=\mbox{exp}(a_pt^p);\;(a_p>0, 0<p\leq 1).$ Inflation is necessary but not
sufficient $(t_{0}^{crit}>t^{inf})$ for $0<p<1.$
\vskip 12pt
\hspace*{-6mm}Fig.2. Cosmic string nucleation $(t\leq t_{0}^{crit})$ near the
inflationary boundary $t<t^{inf}$ in cosmologies with scale factors
$a(t)=\mbox{exp}(a_pt^p);\;(a_p<0, p<0).$ Inflation is necessary but not
sufficient $(t_{0}^{crit}<t^{inf})$.
\vskip 12pt
\hspace*{-6mm}Fig.3. A two phase cosmological model of an expanding
universe with scale factor
$a(t)=\mbox{exp}[-\mbox{exp}(-t)];\;\;t\in\;]-\infty,\infty[\;.$ The
inflationary era
$t\in\;]-\infty,0[\;$ is followed by a flat Minkowski phase
$t>>0.$ Circular cosmic strings may nucleate for $t\leq
t_0^{crit}=-\log(3+2\sqrt{2})$.
\vskip 12pt
\hspace*{-6mm}Fig.4. Circular string evolution near the inflationary
boundary $t<t^{inf}=0$ in a de Sitter-Minkowski expanding toy
model defined by eq.(3.13).
\end{document}